\documentclass[lettersize,journal]{IEEEtran}
\usepackage{amsmath,amsfonts}
\usepackage{algorithmic}
\usepackage{algorithm}
\usepackage{array}
\usepackage{textcomp}
\usepackage{stfloats}
\usepackage{url}
\usepackage{verbatim}
\usepackage{graphicx}
\usepackage{subfigure}
\usepackage{cite}
\usepackage{multirow}
\usepackage{amsmath,amssymb,bm}

\makeatletter
\newcommand*\bigcdot{\mathpalette\bigcdot@{.5}}
\newcommand*\bigcdot@[2]{\mathbin{\vcenter{\hbox{\scalebox{#2}{$\m@th#1\bullet$}}}}}
\makeatother
\begin{document}
\title{Shift-MoE-Based DJSCC for CSI Feedback in Multi-User Pinching-Antenna Systems}
\author{Jian Zou, Yifan Lian, Yongsheng Liang, Fanyang Meng, Wenwu Xie, Liang Yang, and Jian Xiao
\thanks{This work was supported in part by the National Natural Science Foundation of China under Grant 62571160, in part by the Engineering Technology R\&D Center of Guangdong Provincial Universities 2024GCZX004, and in part by the Natural Science Foundation of Hunan Province under Grant 2023JJ50045, Grant 2025JJ70287, Grant 2024JJ7218, and Grant 2024JJ7219. (Corresponding author: Yongsheng Liang.)}
\thanks{Jian Zou is with the College of Applied Technology, Shenzhen University, Shenzhen 518060, China (e-mail: zoujian250@gmail.com). }
\thanks{Yifan Lian is with the College of Electronics and Information Engineering, Shenzhen University, Shenzhen 518060, China (e-mail: yifanll@outlook.com).}
\thanks{Yongsheng Liang is also with the School of Artificial Intelligence, Shenzhen Technology University, Shenzhen 518118, China, and also with the College of Electronics and Information Engineering, Shenzhen University, Shenzhen 518060, China (e-mail: liangys@szu.edu.cn).}
\thanks{Fanyang Meng is with the Research Center of Networks and Communications, Pengcheng Laboratory, Shenzhen 518055, China (e-mail: mengfy@pcl.ac.cn).}
\thanks{Wenwu Xie is with the School of Information Science and Engineering, Hunan Institute of Science and Technology, Yueyang 414006, China (e-mail: gavinxie@hnist.edu.cn).}
\thanks{Liang Yang is with the College of Computer Science and Electronic Engineering, Hunan University, Changsha 410082, China, and also with the School of Communication and Electronic Engineering, Jishou University, Jishou 416000, China (e-mail: liangy@hnu.edu.cn).}
\thanks{Jian Xiao is with the Department of Electronics and Information Engineering, College of Physical Science and Technology, Central China Normal University, Wuhan 430079, China, and also with School of Electrical and Electronics Engineering, Nanyang Technological University (e-mail: jianx@mails.ccnu.edu.cn).}
\thanks{Jian Zou and Yifan Lian contributed equally to this work.}
}

\maketitle

\begin{abstract}
In frequency-division duplexing systems, the performance gains of pinching-antenna systems (PASS) critically depend on accurate channel state information (CSI) at the base station. However, PASS CSI exhibits structured correlations over the waveguide-antenna grid and pronounced heterogeneity across users, making conventional fixed feedback mappings difficult to generalize. To address this challenge, this letter proposes an end-to-end CSI feedback scheme over a noisy uplink feedback link based on deep joint source-channel coding, termed Shift-based Mixture-of-Experts (Shift-MoE). Specifically, Shift-MoE leverages channel-grouped one-step shift operations to capture grid dependencies without global attention, and employs a gated multilayer perceptron mixture-of-experts module to adapt to heterogeneous CSI statistics across users. Numerical results demonstrate that the proposed Shift-MoE consistently outperforms representative learning-based CSI feedback baselines in normalized mean squared error and remains effective under different system parameter settings.

\end{abstract}


\section{Introduction}
\IEEEPARstart{P}{inching}-antenna systems (PASS) have recently attracted increasing attention alongside reconfigurable intelligent surfaces and movable-antenna technologies as promising architectures for reconfigurable propagation, owing to their low power consumption, low cost, and flexible deployment, and offering a new means to manipulate the propagation environment for performance enhancement~\cite{b1,b2,b3}. Existing studies have shown that the achievable gains of PASS critically depend on accurate channel state information (CSI) at the base station (BS) to support downlink precoding and multi-user interference management~\cite{b4}. While many prior works are developed under time-division duplexing settings that leverage channel reciprocity, deploying PASS in practical frequency-division duplexing (FDD) systems is also of strong practical relevance to current commercial networks. In FDD systems, frequency separation prevents exploiting uplink–downlink reciprocity, and accurate CSI feedback becomes a key bottleneck to realizing PASS gains in downlink precoding and multi-user interference mitigation.

Recent deep learning-based CSI feedback methods, exemplified by CsiNet, have shown that compact latent representations can be learned effectively under limited feedback overhead~\cite{b5}. However, most existing schemes essentially formulate CSI feedback as a source compression problem and implicitly assume an error-free feedback link, while practical feedback transmission inevitably suffers from noise and fading, leading to a pronounced source-channel mismatch and performance degradation~\cite{b6,b7}. To address this issue, deep joint source-channel coding (DJSCC) provides an end-to-end framework that jointly optimizes representation extraction and transmission robustness over impaired channels. A seminal work in~\cite{b8} introduced DJSCC for wireless image transmission and demonstrated the advantage of end-to-end learning over noisy channels. Building upon this idea, DJSCC has also been extended to CSI feedback, allowing transmission impairments to be explicitly taken into account during the feedback process~\cite{b9}. Nevertheless, PASS-oriented CSI feedback remains largely unexplored, and in multi-user PASS scenarios, heterogeneous CSI statistics and PASS-specific spatial correlations make a single fixed feedback mapping difficult to generalize across users.

Motivated by the above, this letter studies CSI feedback for multi-user PASS and proposes a Shift-MoE-based end-to-end feedback framework. Specifically, the proposed design integrates a shift-based interaction mechanism to capture structured correlations along the waveguide and antenna dimensions with low complexity, and a gated mixture-of-experts (MoE) module to enhance representation learning under heterogeneous user CSI distributions. Within a DJSCC framework, the encoder and decoder are jointly optimized over the noisy feedback channel to improve robustness against transmission impairments. Numerical results validate the effectiveness of the proposed method in achieving superior CSI reconstruction performance under different user scales, antenna configurations, and compression ratios.

\begin{figure}[!t]
\centering
\includegraphics[width=3.5in]{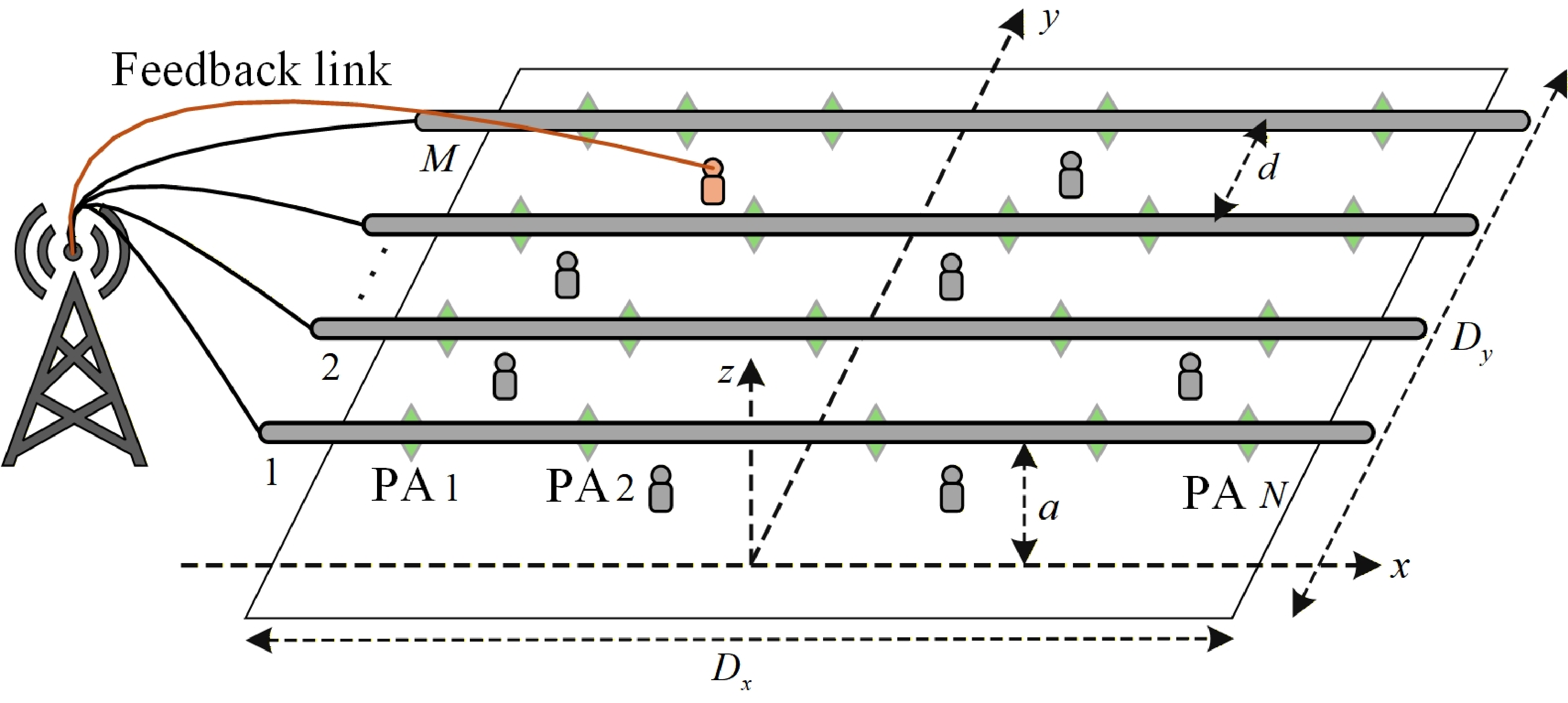}
\caption{Pinching antennas assisted multi-user systems.}
\label{fig1}
\end{figure}

\section{System Model and Problem Formulation}
\subsection{System Model}
As shown in Fig.~\ref{fig1}, we consider a downlink multi-user scenario where the BS serves $K$ users via a PASS consisting of passive dielectric waveguides and pinching antennas (PAs)~\cite{b11}. The PASS comprises $M$ parallel dielectric waveguides deployed at a height $a$ above the user plane. These waveguides are parallel to the $x$-axis and uniformly spaced along the $y$-axis with inter-waveguide distance $d$, where the $m$-th waveguide is located at $y_m=(m-1)d$, $m=1,2,\ldots,M$. Each waveguide is equipped with $N$ fixed PAs placed along the $x$-axis at positions $l_{m,n}=(n-1)\Delta l$, $n=1,2,\ldots,N$, where $\Delta l$ denotes the PA spacing. The $k$-th user is located at $\mathbf{u}_k=(x_k,y_k)$ on the user plane. 
We adopt a geometric multipath channel model consisting of a LoS component and $P$ scattered (NLoS) components~\cite{b10}. Specifically, the PA-level channel coefficient between the $n$-th PA on waveguide $m$ and user $k$ is modeled as
\begin{equation}
h_{k,m,n}=h^{\mathrm{LoS}}_{k,m,n}+h^{\mathrm{NLoS}}_{k,m,n}.
\end{equation}

The propagation distance between the $n$-th PA on waveguide $m$ and user $k$ is denoted by
\begin{equation}
D_{k,m,n}=\sqrt{(l_{m,n}-x_k)^2+(y_m-y_k)^2+a^2},
\end{equation}
and the corresponding LoS channel coefficient is modeled as
\begin{equation}
h^{\mathrm{LoS}}_{k,m,n}
=\frac{\xi}{D_{k,m,n}} e^{(-j\kappa D_{k,m,n})},
\end{equation}
where $\xi=\lambda/(4\pi)$ is the free-space constant, $\lambda=c/f$ is the wavelength with $c$ denoting the speed of light and $f$ the carrier frequency,  and $\kappa=2\pi/\lambda$.

To account for scattering, let the $p$-th scatterer be located at $(x_{s,p},y_{s,p})$ with height $a_s$. The PA-to-scatterer distance and scatterer-to-user distance are given by
\begin{equation}
\begin{aligned}
D_{p,m,n} &= \sqrt{(l_{m,n}-x_{s,p})^2+(y_m-y_{s,p})^2+(a-a_s)^2},\\
D_{k,p}   &= \sqrt{(x_{s,p}-x_k)^2+(y_{s,p}-y_k)^2+a_s^2}.
\end{aligned}
\end{equation}

The NLoS component is then expressed as a superposition of $P$ scattered paths
\begin{equation}
h^{\mathrm{NLoS}}_{k,m,n}
=\frac{\lambda}{(4\pi)^{3/2}}
\sum_{p=1}^{P}
\frac{\beta_{k,p}}{D_{p,m,n}\,D_{k,p}}
e^{(-j\kappa (D_{p,m,n}+D_{k,p}))},
\end{equation}
where $\beta_{k,p}$ denotes the complex gain of the $p$-th scattered path associated with user $k$.

Due to signal propagation inside the dielectric waveguide, an additional phase shift is introduced before radiation at each PA. This effect is captured by the waveguide radiation vector
\begin{equation}
\mathbf{e}_m=\left[p_{m,1}e^{\left(-j \kappa i_{ref} l_{m,1}\right)},\ldots,p_{m,N}e^{\left(-j \kappa i_{ref} l_{m,N}\right)}\right]^T \in\mathbb{C}^{N\times 1},
\end{equation}
where $p_{m,n}$ denotes the coupling gain from waveguide $m$ to its $n$-th PA. In this work, uniform coupling is assumed, i.e., $p_{m,n}=1/\sqrt{N}$. Then, the effective radiation channel from waveguide $m$ to user $k$ is obtained by coherent combining as
\begin{equation}
g_{k,m}(\mathbf{l}_m)=\mathbf{h}_{k,m}^{T}\mathbf{e}_m=\sum_{n=1}^{N} h_{k,m,n}\, p_{m,n}e^{\left(-j \kappa i_{ref} l_{m,n}\right)},
\end{equation}
where $\boldsymbol{\Lambda }=[\mathbf{l}_1,\ldots,\mathbf{l}_M]\in\mathbb{R}^{N\times M}$ collects the PA position vectors on all waveguides. Then, the effective channel vector from all waveguides to user $k$ is defined as $\mathbf{g}_k(\boldsymbol{\Lambda})=\left[g_{k,1}(\mathbf{l}_1),\, g_{k,2}(\mathbf{l}_2),\,\ldots,\,g_{k,M}(\mathbf{l}_M)\right]^T\in\mathbb{C}^{M\times 1}$. To preserve the two-dimensional structural details across the waveguide and PA dimensions, we define the CSI matrix of user $k$ as $\mathbf{H}_k \in\mathbb{C}^{M\times N}$, where the $(m,n)$-th entry is given by 
\begin{equation}
[\mathbf{H}_k]_{m,n}=h_{k,m,n}\,p_{m,n}e^{\left(-j \kappa i_{ref} l_{m,n}\right)},
\end{equation}
which implies $g_{k,m}(\mathbf{l}_m)=\sum_{n=1}^{N}[\mathbf{H}_k]_{m,n}$.

\subsection{Problem Formulation}
We investigate CSI feedback for an FDD multi-user PASS downlink. For each user $k$, the CSI to be fed back is the complex matrix $\mathbf{H}_k \in\mathbb{C}^{M\times N}$. In this work, PASS is used for downlink data transmission, while CSI feedback is conveyed from the UE to the BS via a conventional uplink control/feedback channel. Due to limited feedback overhead and an impaired uplink feedback link, we adopt an end-to-end DJSCC strategy that directly maps $\mathbf{H}_k$ to feedback-channel symbols and optimizes the reconstruction quality over the feedback link. Specifically, we concatenate the real and imaginary parts of $\mathbf{H}_k$ along the last dimension as an $M\times N\times 2$ real-valued input. The encoder ${f_1}(\bigcdot, {{\boldsymbol \theta}_1})$ outputs a real codeword
\begin{equation}
\mathbf{Z}_k=f_{1}(\mathbf{H}_k,{\boldsymbol \theta}_1)\in\mathbb{R}^{L\times 1},
\end{equation}
where $L$ denotes the network output dimension and $L_{c}={L}/{2}$. We interpret $\mathbf{Z}_k$ as two length-$L_{c}$ I/Q sequences to form a complex symbol $\mathbf{s}_k \in\mathbb{C}^{L_{{c}}\times 1}$, which is normalized to satisfy unit average transmit power per block. The compression ratio based on real dimensions is $\eta \triangleq \frac{L}{2MN}$.

The uplink feedback link is modeled as a quasi-static flat Rician fading channel with additive white Gaussian noise (AWGN). Over one feedback block of length $L_{{c}}$, the fading coefficient  $\mathbf{h}_k^{{fb}}$ remains constant, and the received signal at the BS is
\begin{equation}
\mathbf{y}_k^{{fb}} = \mathbf{h}_k^{{fb}}\mathbf{s}_k+\mathbf{n}_k^{{fb}},
\end{equation}
where $\mathbf{n}_k^{{fb}}\sim\mathcal{CN}(\mathbf{0},\sigma_{{fb}}^2\mathbf{I}_{L_{{c}}})$.
The BS obtains $\hat {\mathbf{h}}_k^{{fb}}$ via pilot-aided channel estimation and applies one-tap equalization $\tilde{\mathbf{s}}_k=\mathbf{y}_k^{{fb}}/\hat {\mathbf{h}}_k^{{fb}}$. To focus on the end-to-end DJSCC mapping, we use an idealized receiver-side channel knowledge setting in simulations by taking $\hat {\mathbf{h}}_k^{{fb}} = \mathbf{h}_k^{{fb}}$, which yields $\tilde {\mathbf{s}}_k = \mathbf{s}_k + \mathbf{n}_k^{{fb}}/ \mathbf{h}_k^{{fb}}$.

Given $\tilde{\mathbf{s}}_k$, the decoder $f_2(\cdot;\boldsymbol{\theta}_2)$ reconstructs the CSI and outputs the estimate $\hat{\mathbf{H}}_k\in\mathbb{C}^{M\times N}$ as
\begin{equation}
\hat{\mathbf{H}}_k = f_2(\tilde{\mathbf{s}}_k;\boldsymbol{\theta}_2).
\end{equation}

The encoder and decoder parameters $(\boldsymbol{\theta}_1,\boldsymbol{\theta}_2)$ are optimized end-to-end by minimizing the expected reconstruction error over the CSI distribution and the feedback channel
\begin{equation}
(\boldsymbol{\theta}_1^\star,\boldsymbol{\theta}_2^\star)=
\arg\min_{\boldsymbol{\theta}_1,\boldsymbol{\theta}_2}
\mathbb{E}\!\left[\|\mathbf{H}_k-\hat{\mathbf{H}}_k\|_F^2\right],
\end{equation}
where $\|\cdot\|_F$ denotes the Frobenius norm.

\begin{figure*}[!t]
\centering
\includegraphics[width=5.5in]{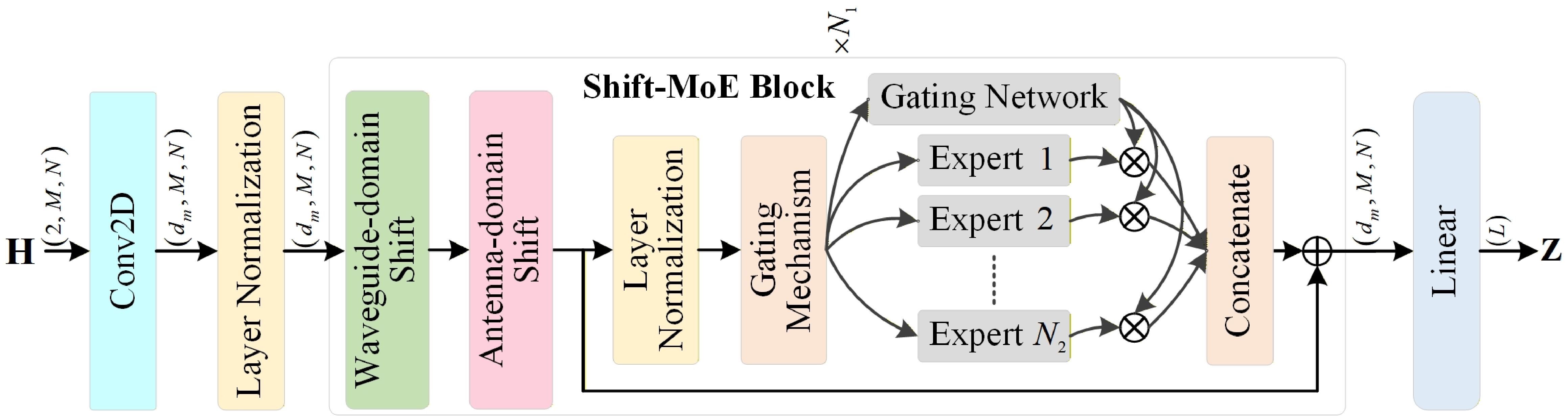}
\caption{Proposed Shift-MoE encoder architecture.}
\label{fig2}
\end{figure*}

\section{Shift-MoE: CSI Feedback for Multi-User Pinching-Antenna Systems}

The CSI dataset is generated based on the considered PASS geometry and the adopted channel model. Specifically, we repeatedly sample user locations $\mathbf{u}_k=(x_k,y_k)$ uniformly in the considered region and compute the PA-level CSI matrix $\mathbf{H}_k\in\mathbb{C}^{M\times N}$. For neural processing, we concatenate $\Re\{\mathbf{H}_k\}$ and $\Im\{\mathbf{H}_k\}$ along the last dimension to obtain an $M\times N\times 2$ real-valued tensor, which is used as the network input and target. 

Motivated by the fact that multi-user PASS CSI exhibits structured spatial correlations along both the waveguide and PA dimensions, while its statistics may vary across users and operating conditions, we propose a Shift-MoE-based CSI feedback framework, as illustrated in Fig.~\ref{fig2}. The proposed design targets two requirements: \emph{i)} capturing the $M\times N$ spatial dependencies without global attention, and \emph{ii)} enhancing representation flexibility through input-adaptive expert modeling. The overall architecture follows a symmetric encoder–decoder structure: the encoder maps the real/imaginary concatenated CSI to a compact latent code for uplink feedback, and the decoder mirrors the encoder to reconstruct the CSI at the BS.

\textit{1) CSI representation and projection:}
We represent the input CSI by concatenating its real and imaginary parts and arranging them into a two-channel tensor $\mathbf{H}\in\mathbb{R}^{2\times M\times N}$. A $1\times1$ Conv2D layer projects $\mathbf{H}$ to a feature tensor $\mathbf{F}_0\in\mathbb{R}^{d_m\times M\times N}$, followed by Layer Normalization (LN). The resulting features are then processed by $N_1$ cascaded Shift-MoE blocks.

\textit{2) Shift mechanism:}
In each Shift-MoE block, the Shift module performs deterministic one-step feature shifts on channel groups to enable local spatial interaction~\cite{b12,b13}. Let $\mathbf{F}\in\mathbb{R}^{d_m\times M\times N}$ denote the block input and let $g=\left\lfloor d_m/n_{div}\right\rfloor$ be the group width under $n_{div}$ channel groups. Using a slicing notation to describe the channel-group shifts, the shifted feature $\mathbf{F}^{{sh}}=\mathcal{S}(\mathbf{F})$ is constructed by shifting the first four channel groups along four directions, while keeping the remaining channels unchanged
\begin{equation}
\begin{aligned}
\mathbf{F}^{{sh}}_{[:,0:g,:,0:N-1]} &= \mathbf{F}_{[:,0:g,:,1:N]},\\
\mathbf{F}^{{sh}}_{[:,g:2g,:,1:N]} &= \mathbf{F}_{[:,g:2g,:,0:N-1]},\\
\mathbf{F}^{{sh}}_{[:,2g:3g,0:M-1,:]} &= \mathbf{F}_{[:,2g:3g,1:M,:]},\\
\mathbf{F}^{{sh}}_{[:,3g:4g,1:M,:]} &= \mathbf{F}_{[:,3g:4g,0:M-1,:]},\\
\mathbf{F}^{{sh}}_{[:,4g:d_m,:,:]} &= \mathbf{F}_{[:,4g:d_m,:,:]}.
\end{aligned}
\end{equation}

The first two equations correspond to shifts along the PA dimension, and the next two correspond to shifts along the waveguide dimension. After shifting, LN is applied. The normalized feature map is then flattened by vectorizing the $M \times N$ grid into a token representation $\mathbf{T}\in\mathbb{R}^{MN\times d_m}$ for expert processing.

\textit{3) MLP-based mixture-of-experts:}
To improve modeling flexibility under heterogeneous CSI statistics, we introduce an MLP-based MoE module with gating control~\cite{b14}. Before routing to experts, a token-wise gating mechanism modulates $\mathbf{T}$ through a two-layer perceptron. Specifically, the gating mask is computed as
\begin{equation}
\mathbf{G}=\sigma\!\big(\mathbf{W}_{g,2}\,\rho(\mathbf{W}_{g,1}\mathbf{T})\big)\in\mathbb{R}^{MN\times d_m},
\end{equation}
where $\rho(\cdot)$ denotes the ReLU activation, and $\sigma(\cdot)$ denotes the Sigmoid activation. The weight matrices satisfy $\mathbf{W}_{g,1}\in\mathbb{R}^{d_m\times \alpha_1 d_m}$ and $\mathbf{W}_{g,2}\in\mathbb{R}^{\alpha_1 d_m\times d_m}$, where $\alpha_1$ controls the hidden expansion of the gating mechanism. The gated tokens are obtained by element-wise suppression
\begin{equation}
\mathbf{T}_g=\mathbf{G}\odot\mathbf{T}\in\mathbb{R}^{MN\times d_m},
\end{equation}
where $\odot$ denotes element-wise multiplication.

We construct $N_2$ expert MLPs $\{\mathcal{E}_n(\cdot)\}_{n=1}^{N_2}$, each implemented as a two-layer feed-forward network with expansion controlled by $\alpha_2$. The $n$-th expert is given by
\begin{equation}
\mathcal{E}_n(\mathbf{T}_g)=\mathbf{W}^{(n)}_{2}\,\phi\!\left(\mathbf{W}^{(n)}_{1}\mathbf{T}_g\right),
\end{equation}
where $n=1,\ldots,N_2$, $\phi(\cdot)$ denotes the GELU activation, $\mathbf{W}^{(n)}_{1}\in\mathbb{R}^{d_m\times \alpha_2 d_m}$ and $\mathbf{W}^{(n)}_{2}\in\mathbb{R}^{\alpha_2 d_m\times d_m}$, and $\alpha_2$ is the expert expansion factor. To adaptively weight experts, we compute a pooled descriptor
\begin{equation}
\bar{\mathbf{t}}=\frac{1}{MN}\sum_{i=1}^{MN}\mathbf{T}_g[i,:]\in\mathbb{R}^{d_m}.
\end{equation}

This descriptor is fed to a gating network consisting of a linear layer and a softmax function. The resulting expert-weight vector is $\mathbf{w}=\mathrm{Softmax}\!\left(\mathbf{W}_{r}\bar{\mathbf{t}}\right)\in\mathbb{R}^{N_2}$, where $\mathbf{W}_{r}\in\mathbb{R}^{N_2\times d_m}$. The MoE output is obtained via weighted aggregation of expert outputs
\begin{equation}
\mathbf{T}_{{moe}}=\sum_{n=1}^{N_2} w_n\,\mathcal{E}_n(\mathbf{T}_g)\in\mathbb{R}^{MN\times d_m}.
\end{equation}

The aggregated tokens are then mapped back to the spatial feature tensor $\mathbf{F}_{{moe}}\in\mathbb{R}^{d_m\times M\times N}$ and combined with the shortcut connection to form the block output $\mathbf{F}_{{out}}=\mathbf{F}^{{sh}}\oplus\mathbf{F}_{{moe}}$, where $\oplus$ denotes element-wise addition. Stacking $N_1$ such blocks enables efficient spatial correlation modeling via shift-based interaction and adaptive nonlinear transformation via gated expert combination.

\textit{4) Symmetric decoder:}
The decoder mirrors the encoder. It first expands the equalized latent code into a spatial feature tensor through a linear layer, applies the same stack of Shift-MoE blocks to progressively recover spatial structures, and finally maps the decoded features back to a two-channel output $\hat{\mathbf{H}}\in\mathbb{R}^{2\times M\times N}$, which is reassembled into the complex CSI estimate.

\section{Numerical Results}

This section validates the performance of the proposed Shift-MoE CSI feedback framework through numerical results. The system parameters, model configurations, and training settings adopted in our simulations are summarized in Table~\ref{Table I}. We evaluate the CSI reconstruction performance using the NMSE, defined as
\begin{equation}
\mathrm{NMSE}= \mathbb{E}\!\left[\frac{\|\mathbf{H}-\hat{\mathbf{H}}\|_F^2}{\|\mathbf{H}\|_F^2}\right],
\end{equation}
where $\mathbf{H}$ and $\hat{\mathbf{H}}$ denote the true and reconstructed CSI matrices, respectively.

\begin{table}[!t]
\begin{center}
\tabcolsep=0.3cm
\renewcommand\arraystretch{1.3}
\caption{System Parameters, Model Configurations, and Training Settings}
\label{Table I}
\begin{tabular}{ c | c  c }
\hline
\hline
\textbf{Parameter} & \textbf{Symbol} & \textbf{Value}\\
\hline 
Rectangular region & $D_x$, $D_y$ & $50~\mathrm{m}$, $6~\mathrm{m}$\\
Waveguides height & $a$ & $5~\mathrm{m}$\\
Inter-waveguide spacing & $d$ & $D_y/(M-1)$ \\
Speed of light& $c$ & $3.0\times 10^{8}~\mathrm{m/s}$\\
Carrier frequency & $f$ & $28~\mathrm{GHz}$ \\
Refractive index & $i_{ref}$ & 1.44\\
Number of scatterers (paths) & $P$ & 3\\
Channel group number & $n_{div}$ & 12\\  
Embedding dimensions & $d_{m}$ & 48\\  
Shift-MoE blocks, experts & $N_1$, $N_2$ & 6, 5\\
Expansion factor & $\alpha _{1}$, $\alpha _{2}$ & 2, 2\\  
Learning rate & -- & $10^{-4}$\\
Batch size, epochs & -- & $32$, $300$\\
\hline 
\hline
\end{tabular}
\end{center}
\end{table}

\begin{figure}[!t]
\centering
\includegraphics[width=2.8in]{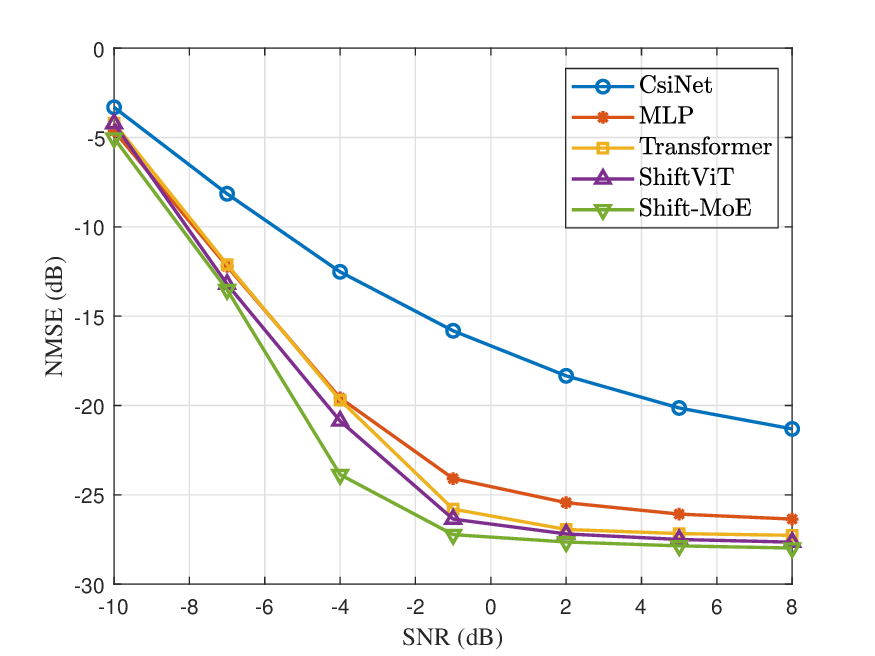}
\caption{NMSE performance of different methods.}
\label{fig3}
\end{figure}

\begin{figure*}[!t]
	\begin{minipage}[t]{0.32\linewidth}
		\centering
		\subfigure[Different numbers of users.]{
			\includegraphics[width=1\linewidth]{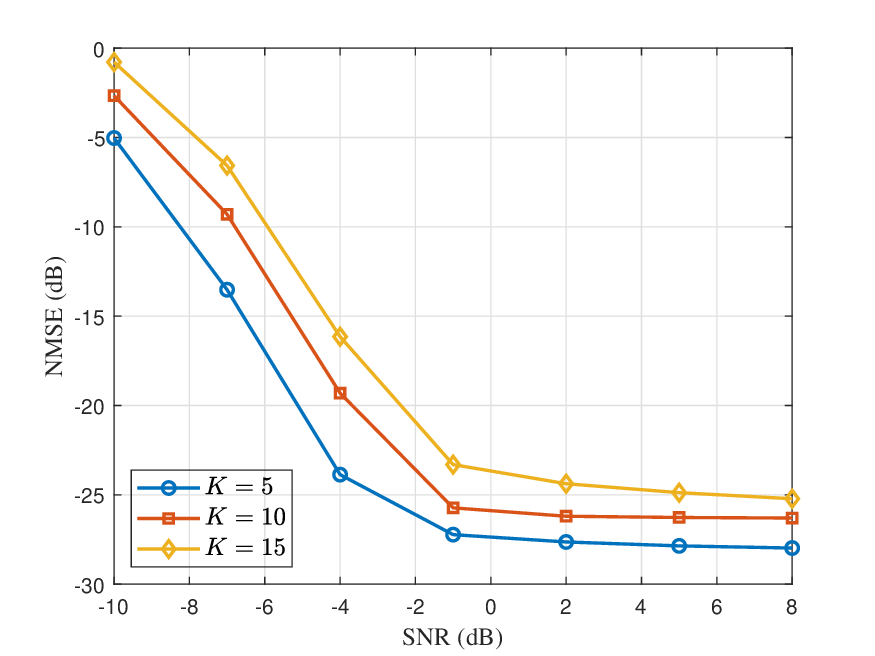}
		}
	\end{minipage}
	\hfill
	\begin{minipage}[t]{0.32\linewidth}
		\centering
		\subfigure[Different compression ratios.]{
			\includegraphics[width=1\linewidth]{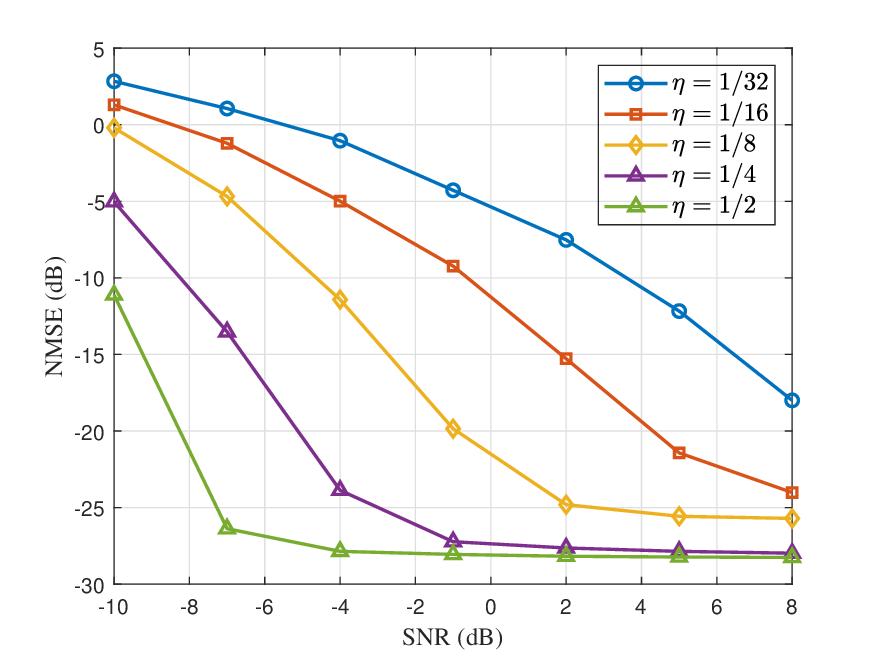}
		}
	\end{minipage}
	\hfill
	\begin{minipage}[t]{0.32\linewidth}
		\centering
		\subfigure[Different antenna configurations.]{
			\includegraphics[width=1\linewidth]{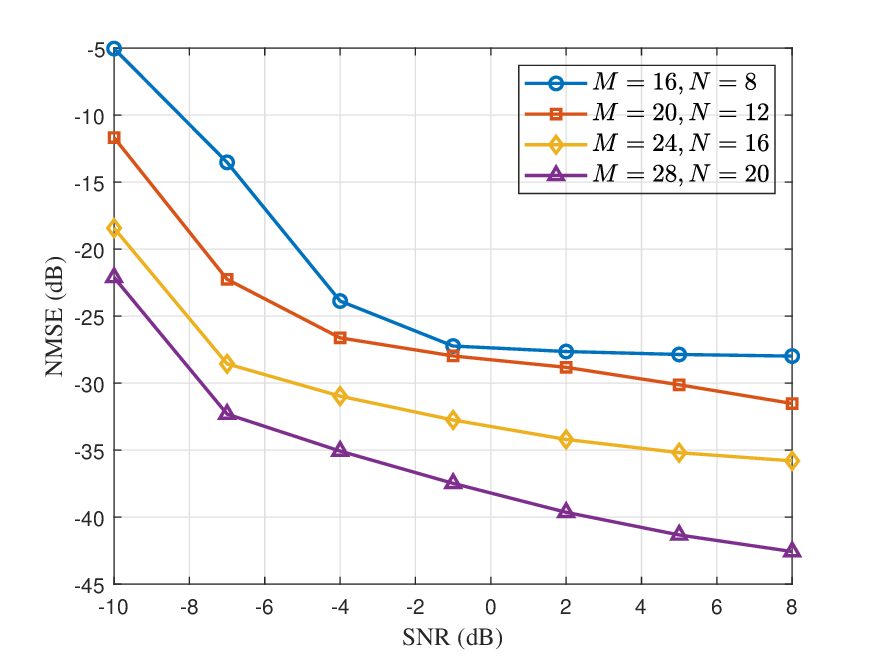}
		}
	\end{minipage}
	\caption{NMSE performance of the proposed Shift-MoE under various parameter settings.}
    \label{fig4}
\end{figure*}

Fig.~\ref{fig3} compares the NMSE performance versus SNR under the configuration of $K=5$, $M=16$, $N=8$, and $\eta=1/2$. The compared methods include CsiNet~\cite{b4}, an MLP baseline without the shift module, Transformer-based schemes~\cite{b7,b15}, as well as the proposed ShiftViT without MoE and Shift-MoE. As SNR increases, the NMSE of all methods decreases due to reduced feedback-link distortions. Compared with the MLP baseline, ShiftViT achieves better performance in the medium-to-high SNR regime, indicating that the shift mechanism can effectively capture the structured correlations of CSI along the waveguide and PA dimensions, thereby improving representation efficiency and reconstruction accuracy. Furthermore, incorporating MoE on top of ShiftViT brings additional NMSE improvements, indicating that expert routing better accommodates heterogeneous CSI statistics in multi-user scenarios. From a complexity perspective, with the sequence length $MN$ and MLP expansion factor $\alpha_{1}$, a Transformer block has complexity $\mathcal{O}\big((MN)^2 d_m+\alpha_{1}MN,d_m^2\big)$, where the quadratic term comes from MHSA. Replacing MHSA with channel-grouped one-step shifts removes the attention quadratic term, and a ShiftViT block is dominated by the MLP term, scaling as $\mathcal{O}\big(\alpha_{1}MN,d_m^2\big)$.

\begin{figure}[!t]
\centering
\includegraphics[width=2.85in]{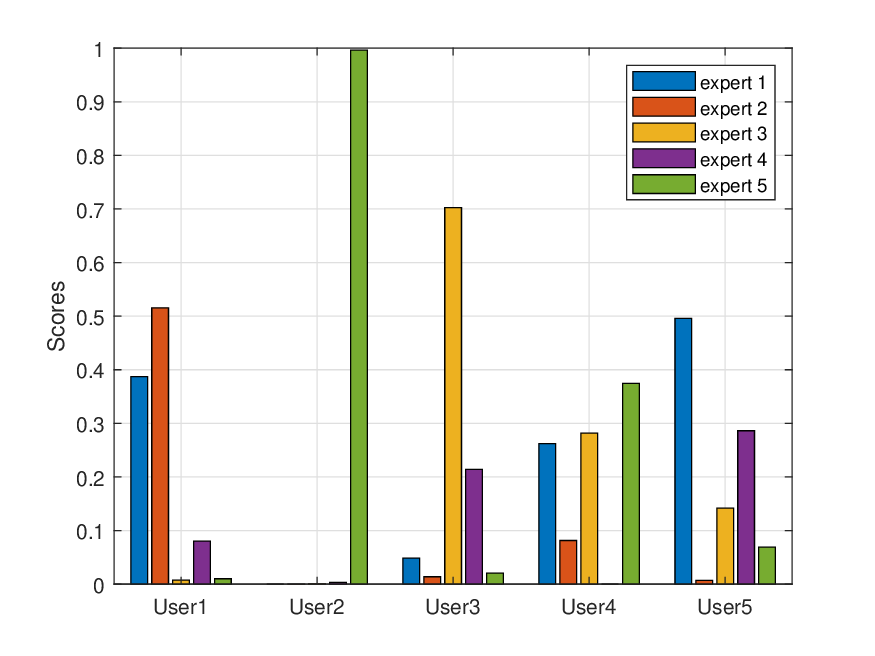}
\caption{The proposed Shift-MoE gating weights for different users.}
\label{fig5}
\end{figure}
Fig.~\ref{fig4}(a) shows the NMSE performance under different numbers of users with $M=16$, $N=8$, and $\eta=1/2$. As $K$ increases, the NMSE degrades over the considered SNR range, because a larger user set intensifies CSI heterogeneity and makes the feedback representation more difficult to learn. The degradation is more evident at medium-to-high SNR, where reconstruction is less noise-limited and more dependent on the model's ability to capture user-dependent structures.
Fig.~\ref{fig4}(b) depicts the effect of the compression ratio $\eta$ under $K=5$, $M=16$, and $N=8$. A larger $\eta$ consistently improves reconstruction accuracy by providing a higher-dimensional feedback representation and reducing information loss. The gain is more pronounced at moderate-to-high SNR, indicating that representation capacity becomes the dominant factor once feedback-channel distortions are sufficiently suppressed.
Fig.~\ref{fig4}(c) evaluates different antenna configurations under $K=5$ and $\eta=1/2$. Increasing $(M,N)$ leads to lower NMSE, suggesting that the proposed design can effectively exploit the richer spatial structure of larger waveguide-PA grids. Overall, the results confirm that Shift-MoE maintains favorable scaling behavior with respect to user scale, feedback rate, and PASS dimensions.

Fig.~\ref{fig5} visualizes the expert gating weights produced by Shift-MoE for different users. The weight patterns vary notably across users, indicating that the MoE module performs input-adaptive routing rather than applying a fixed mapping to all CSI realizations. This adaptive expert fusion aligns well with the multi-user PASS setting, where CSI statistics can differ across users and operating conditions.

\section{Conclusion}
To the best of our knowledge, this letter is among the first to systematically investigate CSI feedback for multi-user PASS under FDD operation. To address the structured waveguide–PA correlations and the pronounced heterogeneity of CSI statistics across users, we develop an end-to-end DJSCC-based feedback framework with the proposed Shift-MoE architecture. Numerical results demonstrate that combining structure-aware modeling with adaptive expert learning effectively improves CSI feedback performance in multi-user PASS. Future work will consider more practical settings such as imperfect feedback-channel knowledge and pursue joint optimization with beamforming and resource allocation.

\end{document}